\documentclass[twocolumn,floatfix,amsmath,amssymb]{revtex4}
\usepackage{graphicx}
\usepackage{epsfig}
\usepackage{epsf}
\usepackage{grffile}

\newcommand\beq{\begin{equation}}
\newcommand\eeq{\end{equation}}
\newcommand\bea{\begin{eqnarray}}
\newcommand\eea{\end{eqnarray}}
\newcommand{\nonum}{\nonumber}
\begin{document}

\title{\bf Collapse and revival of entanglement of two-qubit in 
superconducting quantum dot lattice with  
magnetic flux and inhomogeneous gate voltage }

\author{\bf Sujit Sarkar}
\address{\it PoornaPrajna Institute of Scientific Research,
4 Sadashivanagar, Bangalore 5600 80, India.\\    
}
\date{\today}

\begin{abstract}
We study the entanglement of a two-qubit system in a superconducting
quantum dot (SQD) lattice in the presence of 
magnetic flux and gate voltage inhomogeneity. 
We observe a universal feature for 
the half-integer magnetic flux quantum which 
completely washes out the entanglement of the system both at
zero and finite temperature. 
We observe that the 
ground state is always in a maximally entangled Bell state 
when there is no inhomogeneity in gate voltage in the 
superconducting quantum
dot lattice. We find
an important constraint in magnetic flux for ground state
entanglement. We also observe few behavior of entanglement at
finite temperature is in contrast with the zero temperature behavior.\\ 


\end{abstract}

\maketitle


\section{ Introduction}
Recent years have witnessed a surge in research activity involving 
a close interaction between the subjects of quantum information
science and many body condensed matter physics \cite{ami,lew,fazio,ind}
. The key concept in
quantum information is that of entanglement, a truly unique feature
of quantum mechanical systems. Entanglement implies non-local
correlations between quantum particles which do not have a classical
counterpart. Entanglement measures provide an additional characterization
of the many-body state in condensed matter. Let us consider
the state of an interacting spin system as an example. Since the ground state
wave function changes qualitatively in a quantum phase transition,
it is of significant interest to probe how the genuine quantum aspects
of the wave function, namely entanglement,
changes as the transition point is traversed.
The question that arises is whether the entanglement of the
quantum system extends over the macroscopic distances as ordinary
correlation do. In recent years, quantum phase transitions have been
extensively studied in spin systems using well known quantum information
theoretic measures like entanglement, fidelity, reduced fidelity, fidelity
susceptibility. The fidelity typically drops in an abrupt manner at a
quantum critical point indicating a dramatic change in the nature of
the ground state wave function \cite{gu,quan1,zana,zana2,zhou,chen}.\\
We have been motivated by the extensive studies of entanglement physics
and interesting results in spin 
system \cite{nil,wang,kamta,woo,woo2,sun,yeo,dvk,zhang,bose,datta,loss,feng,yang1}
. We decided to apply this concept
of study in the different disciplines of quantum condensed matter quantum 
many-body system. 
Here we study the
entanglement physics of two qubit superconducting quantum 
dot lattice. Before we
proceed further, we would like to discuss the basic aspects of
a superconducting quantum dot lattice: Superconducting quantum dot
lattice consists of array of superconducting grains. 
The superconducting grains are
of nanoscale size, and different states of the superconducting
grain are controlled by the ratio between charging energy and
Josephson energy, while the average charge of the dot is controlled
by the gate voltage. These are similar to the metallic and semiconductor
quantum dot, where the different phases of the 
system depends on the ratio of Coulomb charging energy to
the kinetic energy. Therefore the system can be described as a superconducting 
quantum dot (SQD) and the lattice as a superconducting quantum dot 
\cite{ss,ss1,ss2,chow,lar,jae,orr} lattice. \\
One can consider the Cooper pair of SQD as a charged boson.
The Physics of SQD can therefore be described in terms of
interacting bosons. Bosonic physics is more interesting
and hard to understand than the fermionic physics. 
One can understand the starting point of fermionic physics
from the
standpoint of independent electron approximation, at least
in higher dimensions, whereas to understand the interacting
bosonic system, one has to introduce the interaction from
the very beginning. 
At the same time quantum phase diagram of this SQD lattice is
very rich with different quantum phases.
Therefore the study of the entanglement physics
for the SQD lattice 
for both zero and finite temperature is interesting in its 
own right [25-27].\\
It is well known from our previous studies that mesoscopic 
SQUID array can also be treated as the superconducting quantum dot lattice
with modulated Josephson junction [25-26]. The authors of Ref. \cite{chow} have found 
the magnetic flux induced superconducting Coulomb blocked in 
mesoscopic SQUID array and also the magnetic flux induced superconductor-
insulator quantum phase transition. Experimentally and also
theoretically it reveals that the applied magnetic flux has an important
effect in the SQD lattice system. We will see in due course of our
study that the inhomoginity of the gate voltage plays an important
role in entanglement to disentanglement (product state) transition. Therefore 
we are motivated to 
study entanglement physics of SQD lattice in the presence of 
applied magnetic flux and inhomoginity of gate voltage.\\
The plan of the manuscript
is as follows. 
We present the model Hamiltonian
and entanglement physics in section (II) of this manuscript.
We present summary and conclusion in the section (III) of
the manuscript. 
\\
%
%
\section{ Model Hamiltonian for inhomogeneous Superconducting
Quantum Dot Lattice and
the study of Entanglement physics}

{\bf II A. Model Hamiltonian and Ground State Analysis}

At first we write down the model Hamiltonian of 
SQD lattice system with 
Josephson couplings having on-site charging energies and inter-site
interactions in the presence of gate voltage and external magnetic flux. 
We also consider the  
inhomogeneity in the applied gate voltage. 
The Hamiltonian is written
as
\beq
H~=~H_{J1}~+~H_{EC0}~+~H_{EC1}.
\eeq
We recast different parts of the Hamiltonian in quantum
phase model as\\
\begin{center}
$
H_{J1}~=~ -E_{J1} |cos(\pi \frac{\Phi}{{\Phi}_0})| \sum_{i}  
cos ({\phi}_{i+1} -{\phi}_{i}),
$
\end{center}
where 
${\phi}_i $ and $ {\phi}_{i+1} $ are quantal phase of the SQD at the point
i and i+1 respectively. 
as
\begin{center}
$
H_{EC0}~=~ \frac{E_{C0}}{2} \sum_{i} 
{(-i \frac{\partial}{{\partial}{{\phi}_i}} - \frac{N_i}{2})^{2} }
$,
\end{center}
where ${E_{C0}}$ is the on-site charging energy,
Now
$$
H_{EC1}~=~ E_{Z1} \sum_{i} {n_i}~{n_{i+1}},
$$
where
$E_{Z1}$ is the NN charging 
energies between
the dots respectively. 
In the phase representation,  
$(-i \frac{\partial}{{\partial}{{\phi}_i}})$ is the operator
representing the number of Cooper pairs at the ith dot, 
and thus it takes only the integer
values ($n_i$). 
Here, Hamiltonian $H_{EC0}$ accounts for the
influence of gate voltage ($e N \sim V_g$), where  
$e N$ is the average dot charge induced by the gate voltage.
When the
ratio $\frac{E_{J1}}{E_{C0}} \rightarrow 0$, the SQD 
array is in the insulating state having a gap of the width
$\sim {E_{C0}}$, since it costs an energy $\sim E_{C0}$
to change the number of pairs at any dot. The exceptions are the 
discrete points at $N~=~(2n+1)$, where a dot with charge $2ne$
and $2 (n+1) e$ has the same energy because the gate charge 
compensates the charges of extra Cooper pair in the dot.
On this degeneracy point, a small amount of Josephson coupling 
leads the system to the superconducting state.\\
Here we recast our basic Hamiltonians in the spin 
language, where each site of the dot is either empty or singly occupied. 
During this process we follow Ref. \cite{ss} and \cite{lar}.
Now
\begin{center}
$
H_{J1}~=~ -2~E_{J1}  |cos(\pi \frac{\Phi}{{\Phi}_0})| \sum_{i} 
( {S_i}^{\dagger} {S_{i+1}}^{-} + h.c)
$,
\end{center}
and
\begin{center}
$
H_{EC0}~=~ \frac{E_{C0}}{2} \sum_{i} 
{(2 {S_i}^{Z} - h )^{2} }.$
$
H_{EC0}~=~ -2 {E_{C0}} \sum_{i} {h_i}
{S_i}^{Z}.$
\end{center}

Here $h_i = \frac{N_i - 2n - 1}{2} $ allows the tuning of the system around the
degeneracy point by means of gate voltage. We can tune the gate voltage in such
a way that we can generate inhomogeneity in on-site charging energy. Without loss
of generality we can also write the model Hamiltonian as
\beq
H_{EC0}~=~ \sum_{i} (E_{C0} + \delta V_g ) {S_i}^{z} 
+ \sum_{i} (E_{C0} - \delta V_g ) {S_{i+1}}^{z}. 
\eeq 
Where $\delta V_g $ is the variation of gate voltage
around the lattice sites.

$ H_{EC1}~=~2 E_{Z1} \sum_{i} {S_i}^{z}~{S_{i+1}}^{z}.
$ \\
The total Hamiltonian of the system is 
\bea
H & = &  2 E_{J1}  |cos(\pi \frac{\Phi}{{\Phi}_0})| \sum_{i}
( {S_i}^{\dagger} {S_{i+1}}^{-} + h.c) +  2 E_{z1} \sum_{i} {S_i}^{z}~{S_{i+1}}^{z} \nonum\\
& & + \sum_{i} (E_{C0} + \delta V_g ) {S_i}^{z} +  
\sum_{i} (E_{C0} - \delta V_g ) {S_{i+1}}^{z}
\eea
\begin{figure}
\includegraphics[scale=0.55,angle=0]{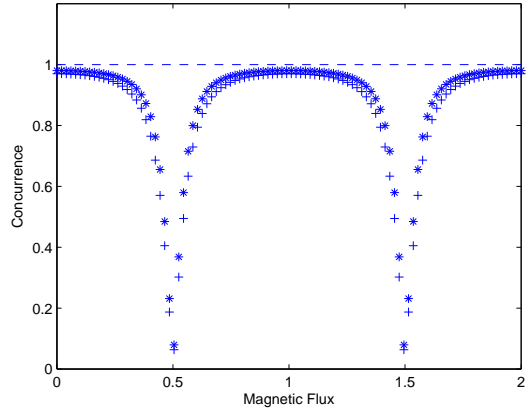}
\caption{Color online, concurrence vs magnetic flux at zero temperature for
different values of $\delta V_g = 0.0, 0.2, 0.5$ for the dashed
bullet and $+$ signs respectively. Here $E_J = 1.0, E_{C0}=1.0$
and $ {E_{Z1} = 0.4}$. }
\label{Fig. 1}
\end{figure}
\begin{figure}
\includegraphics[scale=0.55,angle=0]{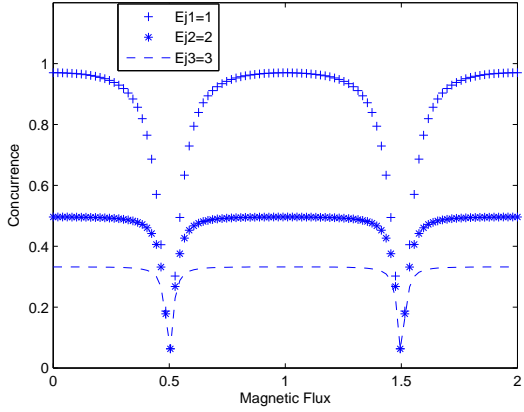}
\caption{Color online, concurrence vs. magnetic flux at zero temperature for different values
of Josephson couplings. Here $E_j = 1.0, \delta V_g =0.5$ and $E_{Z1} = 0.5 $.
}
\label{Fig. 2}
\end{figure}
Now we consider the Hamiltonian for $ N=2 $ case. We would like to
write the Hamiltonian in the standard basis, $ |1,1>,
|1,0>, |0,1>, |0,0> $
\[ H = \left (\begin{array}{cccc}
       E_{Z1} +  E_{C0} & 0  & 0 & 0 \\
       0 &  -  E_{Z1} +  \delta V_g    & B & 0 \\
 0 &  B  & -  E_{Z1} -  \delta V_g & 0 \\
0 & 0 & 0 &  E_{Z1} - E_{C0}  \\ 
        \end{array} \right )  \]
$ B = {2 E_{J1}} |cos(\pi \frac{\Phi}{{\Phi}_0})| $. 
The eigenstates of this two sites Hamiltonian are 
$| { {\psi}_1 } >  = |0,0> $, $ | {{\psi}_1} >  = |1,1> $, 
$ |{ {\psi}_3} > = \sqrt{\frac{1}{1 + {c_1}^2 }} ( {c_1} |1,0> + |0,1> )$,
$ |{ {\psi}_4} > = \sqrt{\frac{1}{1 + {c_2}^2 }} ( {c_2} |1,0> + |0,1> )$.
Where ${c_1} = \frac{ \delta V_g - A}{B}$,
$A = \sqrt{ {\delta V_g}^2 + 4 {E_J}^2 {|cos(\pi \frac{\Phi}{{\Phi}_0})|}^2 }$, 
$ {c_2} = \frac{ \delta V_g + A }{B} $.
$ {E_1} = \frac{1}{2} ( 2 E_{Z1} - 2 E_{C0} ) $, 
$ {E_2} = \frac{1}{2} ( 2 E_{Z1} + 2 E_{C0} ) $,
$ {E_3} = - E_{Z1} - A  $
$ {E_4} = - E_{Z1} + A  $.
If we consider the homogeneous system, i.e.,
there is no variation of gate voltage 
over the lattice sites.
The two states $ | {{\psi}_3}> $ and $ | {{\psi}_4}> $ are the maximally 
entangled Bell states, i.e., $(1/\sqrt{2}) ( |0,1> - |1,0> )$    
, $(1/\sqrt{2}) ( |0,1> + |1,0> )$. As we see from our analytical
expression that ground state depends on the value of $E_{C0} $ 
, $E_{Z1} $ and $A$.
Ground state is in the disentangle state (product state) when the
ground state energy is either $ E_1 $ or $ E_2 $, otherwise the system is
in the entangle state. 
Thus for this superconducting
quantum dot lattice system there is a transition between the disentangle
state to entangle state due to the variation of the system parameters. 
We will see in due course of our study that the
magnetic flux plays an important role in the transition between the
disentangled state to the entangle state. 

{\bf II. B Entanglement Study for Zero and Finite Temperature}\\
Now we calculate the thermal entanglement of 
two arbitrary qubits $(N=2)$
of the superconducting quantum dot lattice.
The density matrix of superconducting quantum dot lattice in equilibrium at temperature,
$ T$ is $ \rho = \frac{1}{Z} exp(-H/{k_B T} ) $, where $H$ is the Hamiltonian of the
system, ${Z}$ is the partion function of the system, $k_B $ is the Boltzmann constant.
We calculate the concurrence to measure the entangle  
of two qubit system of superconducting quantum dot lattice 
following the reference of Wootters' formula [14,15]. The analytical expression for
the concurrence is
\beq
C = max (0, 2 max {\lambda}_i - \sum_{i} {\lambda}_i ) 
\eeq
, where ${\lambda}_i $ is the square roots of the eigenvalues of the
matrix
$ R = \rho ( {{\sigma}_1}^{y} \otimes {{\sigma}_2}^{y}) {\rho}^{*} 
( {{\sigma}_1}^{y} \otimes {{\sigma}_2}^{y}) $.
$ {\rho}^{*} $ is the complex conjugate of ${\rho}$. The system is
maximally entangled when $ C=1$ and the system is in disentangle
state when $C= 0$. 
\\ 
The density matrix of the system is
\[ \rho = \left (\begin{array}{cccc}
       A_1  & 0  & 0 & 0 \\
       0 &  B_2   &  C_2  & 0 \\
 0 & C_2  & C_3 & 0 \\
0 & 0 & 0 & D_4 \\ 
        \end{array} \right ) \]
The square roots of the eigenvalues of the matrix R are
${\lambda}_1 = {\lambda}_2 = \sqrt{A_1 D_4} $,
$ {\lambda}_3 = \sqrt{{B_2 C_3} + C_2 }$ and
$ {\lambda}_4 = \sqrt{{B_2 C_3} - C_2 }$. 
Where
$ A_1 = \frac{1}{Z} e^{-{E_{Z1}}/T} [ cosh( {E_{C0}}/T) 
- \frac{1}{{E_{C0}} T} sinh ({E_{C0}}/T ) $ 
$ D_4 = \frac{1}{Z} e^{-{E_{Z1}}/T} [ cosh( {E_{C0}}/T) 
+ \frac{1}{{E_{C0}} T} sinh ({E_{C0}}/T )] $
$ C_2 =- \frac{B}{ZA} e^{\frac{E_{Z1}}{T}} sinh(A/T) $

$ B_2 =  \frac{1}{Z} e^{{E_{Z1}}/T} [ cosh({A}/T) - 
\frac{\delta V_g}{A} sinh(A/T) ] 
$;
$ C_3 =  \frac{1}{Z} e^{{E_{Z1}}/T} [ cosh({A}/T) + 
\frac{\delta V_g}{A} sinh(A/T) ] 
$. 
In this derivation we consider ${k_B} = 1$.

Finally we obtain the formula for concurrence by using the
relation Eq. 5,
\beq
C = \frac{2}{Z}~e^{E_{z1}/T}~(4 E_J sinh (\frac{A}{T})
   ~ - ~ e^{-2 E_{z1}/T } )
\eeq
Now we calculate, the concurrence of the system by using the above
relation and  the final expression for the concurrence at $T=0 $ is 
\beq
C = |B/A|
\eeq
Now we analyze the above equation of concurrence in different limit.
If we consider the gate voltage in the lattice is homogeneous ,i.e.,
$\delta {V_g} = 0 $ that implies the concurrence
$ C (T=0) = 1 $. The system is always in maximally entangled state.
For this situation magnetic flux has no effect to make a transition from
entangled state to disentangled state.\\
If we consider the presence of inhomogeneous gate voltage, $\delta {V_g} \neq 0$ then
the system has finite concurrence but it has not maximally entangled for any
arbitrary magnetic flux.  We observe an universality that 
at half-integer values of magnetic flux quantum, the
concurrence is zero.\\
The analytical expression which we derive for the concurrence is obtained
satisfies the following relation
\beq
\sqrt{ {\delta V_g}^2 + 4 {E_J}^2 |cos(\frac{\pi \phi}{{\phi}_0})| }^2 \geq { E_{C0} - 2 E_{Z1} }
\eeq
From the analysis of the above equation we get the constraint of magnetic flux that
to obey the concurrence relation system.
\beq
\phi \geq \frac{{\phi}_0}{\pi} cos^{-1}  
\frac{\sqrt{( E_{C0} - 2 E_{Z1} - \delta V_g) }
\sqrt{( E_{C0} - 2 E_{Z1} + \delta V_g})}{2 E_{J} }
\eeq

\begin{figure}
\includegraphics[scale=0.55,angle=0]{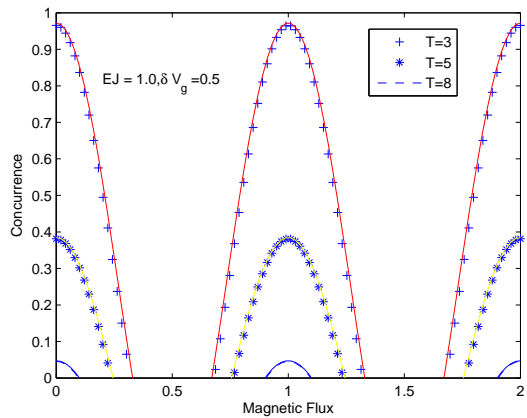}
\caption{Color online, concurrence vs. magnetic flux at different finite temperatures.
Here $E_J = 1.0, \delta V_g =0.5, E_{C0} = 1.0 $ and $E_{z1} =0.5 $.
}
\label{Fig. 3}
\end{figure}
\begin{figure}
\includegraphics[scale=0.55,angle=0]{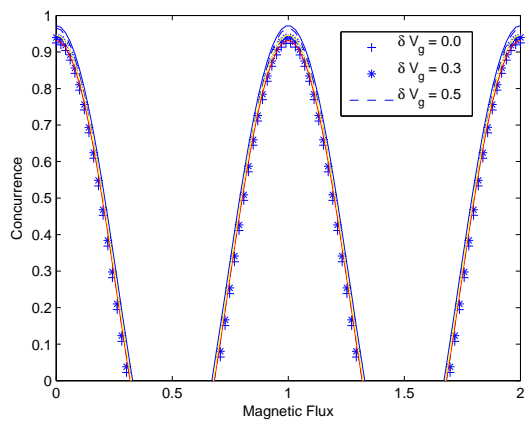}
\caption{Color online, concurrence vs. magnetic flux for different inhomogeneity of gate voltage
in SQD lattice. Here $E_J = 1.0, E_{C0} = 1.0$ and $E_{z1} = 0.5$.
}
\label{Fig. 4}
\end{figure}
Now we present our results and physical explanations.
Fig. 1, shows the variation of concurrence with magnetic flux   
for different
values of $\delta V_g$. We observe from the figure that the
system is maximally entangled Bell  
state when there is no inhomoginity in the applied voltage in the
lattice. 
This maximally entangled Bell state is independent of the
applied magnetic flux.
We also observe from this figure that a finite
inhomogeneity in the applied gate voltage introduces 
a transition from entangled state to disentangled state.
In the entanglement phase system is almost maximally entangled state.\\
Fig. 2 shows the variation of concurrence with magnetic flux for
different values of Josephson couplings. We observe from our study that
concurrence is large for the smaller values of ratio between the
Josephson coupling to on-site Coulomb charging energies. As we have
obtained from our previous studies that this large ratio 
of $\frac{E_J}{E_{C0}}$ favors the Luttinger liquid and 
superconducting phase of the system and the smaller one favors
the Mott insulating density wave phases of the system [25,26]. Thus it is
clear from our study that entanglement of the system is large for
Mott-insulating phase of the system where the Cooper pairs are
localized. But the collapse of entanglement at the half-integer
magnetic flux quantum has the universal feature for all values of
Josephson couplings.\\  
Fig. 3 shows the variation of concurrence with magnetic flux at
different temperatures. It reveals from our study that the 
concurrence is smaller for the higher values of temperature and
at the same times the system is in the disentangled state for
wider region of magnetic flux. The region of disentanglement state
is larger for the higher values of temperature. The  
solid lines in the figures are for the considerations of
superconducting Coulomb blocked effect induced co-tunneling effect
in the system. We observe that co-tunneling effect has not drastic
effect to change  entangle transition of the system at the 
quantitative level. For ground state entanglement,
the effect of co-tunneling effect is not explicit in the
analytical expression (Eq. 8). \\
Fig. 4 shows the variation of concurrence with magnetic flux for
different values of $\delta V_g$. Here we observe that the presence
and absence of gate voltage inhomogeneity has no appreciable for the
transition from entangle state to disentangle state. This behavior
at finite temperature is in contrast with the behavior at zero 
temperature where the system is always entangled state for 
homogeneous gate voltage.\\
  
\section{ Summary and Conclusions}
We have studied the entanglement of a two qubit system of a
superconducting quantum dot lattice. We have observed the magnetic flux
dependent entangled state to disentangled state transition at zero and finite
temperature. We find an important constraint on magnetic flux during the
study of ground state entanglement. We have observed that at zero temperature
that the system is always maximally entangled state when there is no-inhomogeneity
in the gate voltage in the lattice sites whereas the presence and absence
of inhomoginity has no effect at finite temperature. We have found out the universal
feature for half-integer magnetic flux quantum. 

The author would like to acknowledge 
the CCMT of the physics department of IISc for extended facility use.
The author would like to acknowledge Dr. R. Srikanth for reading the
manuscript critically.\\ 

\end{document}